\documentclass[usenatbib]{mn2e}
\usepackage[pdftex]{graphicx}
\usepackage[normalem]{ulem}
\newcommand{\mnras}{MNRAS}

\def\apj{ApJ}
\def\apjl{ApJL}
\def\apjs{ApJS}

\def\aap{A\&A}

\def\mnras{MNRAS}

\title[Stellar Populations in Superclusters of Galaxies]
{Stellar populations in superclusters of galaxies}
\author[M. V. Costa-Duarte L. Sodr\'e Jr. and F. Durret]{M. V.
Costa-Duarte$^{1,2}$\thanks{e-mail:
mvcduarte@astro.iag.usp.br} L. Sodr\'e Jr.$^{1}$ and F. Durret$^{3}$\\
$^{1}$Instituto de Astronomia, Geof\'isica e Ci\^encias Atmosf\'ericas, Universidade de S\~ao Paulo, S\~ao Paulo, Brazil\\
$^{2}$LUTH, Observatoire de Paris, CNRS, Universite Paris Diderot, Place Jules
Janssen, 92190 Meudon, France \\
$^{3}$UPMC Universit\'e Paris 06, UMR~7095, Institut d'Astrophysique de
Paris, 98bis Bd Arago, F-75014, Paris, France}

\begin{document}

\date{}

\pagerange{\pageref{firstpage}--\pageref{lastpage}} \pubyear{2002}

\maketitle

\label{firstpage}

\begin{abstract} 
 A catalogue of superclusters of galaxies is used to investigate the
influence of the supercluster environment on galaxy populations,
considering galaxies brighter than M$_r<$-21+5$\log$ h. 
Empirical spectral synthesis techniques are applied to obtain the stellar
population properties of galaxies which belong to superclusters and
representative
values of stellar population parameters are attributed to each
supercluster. We show that richer superclusters present denser
environments and older stellar populations.
The galaxy populations of
superclusters classified as filaments and pancakes are statistically similar,
indicating that the morphology of
superclusters does not have a significative influence on the stellar
populations. Clusters of galaxies within superclusters
are also examined in order to evaluate the
influence of the supercluster environment on their galaxy properties. 
Our results suggest that the environment affects galaxy properties but its
influence should operate on scales of groups and clusters, more than on
the scale of superclusters.
\end{abstract}

\begin{keywords}
Cosmology: Large-scale structure in the Universe - superclusters of galaxies
- spectral synthesis: galaxies.
\end{keywords}

\section{Introduction}
%
%

Large scale structures are an important tool to study the Universe,
presenting several useful features to constrain the cosmology. Groups
and clusters of galaxies represent two fundamental classes of objects 
in this scenario, providing information on galaxy evolution and structure
formation. Many cluster catalogues have been compiled, presenting 
properties for hundreds or even thousands of objects
\citep{Abelletal1989,Koesteretal2007,Wenetal2009}. 
Beyond group and cluster  scales, the large scale structure is traced by the
largest type of object known: superclusters of galaxies, extending to
tens of megaparsecs and presenting a variety of morphologies. 
%
%
The first galaxy redshift surveys have allowed the compilation of the 
first supercluster catalogues, presenting data for about a hundred structures
\citep{Zuccaetal1993,Einastoetal1994}. Since then, the completion
of new surveys led to a new generation of catalogues. For example,
the Las Campanas Galaxy Redshift survey \citep{Schectmanetla1996} and the
2-degree Field Galaxy Redshift Survey \citep{Collessetal2001}, with tens 
to hundreds of thousands of galaxies, have produced catalogues with several
hundreds of superclusters \citep[][]{Einastoetal2007a}. Nowadays, the Sloan 
Digital Survey \citep[SDSS,][]{Abazajianetal2009} provides 
a database with roughly one million galaxy spectra, making possible a
detailed study of a variety of supercluster properties. A detailed map
of the Universe has been produced, showing, among others, the most prominent
supercluster known, the Sloan Great Wall \citep{Gottetal2005}. Besides,
many efforts are being applied to provide deeper galaxy surveys, such as
zCOSMOS \citep{Lillyetal2007} and GOODS/VIMOS \citep{Popessoetal2009}. Deep
surveys make possible the identification of structures beyond the
local Universe, reaching
z$\sim$1-3 \citep{Tanakaetal2001,GalRubin2004,Kuiperetal2012}.
%
%

We have
compiled a supercluster catalogue using the SDSS/DR7 database
\cite[][hereafter CD11]{Costa-Duarteetal2011}. Our analysis has shown that
superclusters classified as filaments tend to be richer and more luminous. Using
mock lightcones \citep{Crotonetal2006} we also concluded that peculiar
velocities do not represent a significative influence on supercluster
morphologies.
\cite{Einastoetal2011a,Einastoetal2011b} also used the SDSS/DR7 database to 
discuss the relation between morphology and evolution in these structures.
\cite{Luparelloetal2011} presented a detailed study of the future
of superclusters, collapsing into future virialized structures. Recently
\cite{Einastoetal2012} have shown that the cluster population in
superclusters depends on their richness, with richer
clusters being preferentially located in the high density regions of rich 
superclusters. 
%
%

It is well known that  galaxy environment has an important role on 
galaxy evolution, affecting galaxy morphology and stellar
populations \citep{DavisGeller1976,Dressler1980,Whitmoreetal1993}. 
High density environments are usually populated by early-type
galaxies, presenting old stellar populations, while late-type galaxies,
with younger stellar populations, are mostly found
in low density environments. Environmental processes may act even at very
large scales \citep{Moetal1992,Mateusetal2007}. Several
studies have shown this effect by considering the environmental effect on
colors, morphology and luminosity function of galaxies
\citep[see][]{Crotonetal2005,Parketal2007,Carolloetal2012}.
\cite{Einastoetal2007b}
have compared the galaxy population properties of rich and poor superclusters
in the 2dFGRS
by using color index and spectral features,
concluding that richer structures present a higher fraction of
early-type galaxies than the poorer ones. They also show that
main galaxies of rich superclusters are more luminous than the main galaxies in
poor structures. Recently, \cite{Lietzenetal2012} have shown that
groups of galaxies belonging to superclusters evolve faster than
groups in the field or in voids, presenting higher luminosities and higher fractions of
passive galaxies.
%
%

In this paper we investigate the connection between stellar populations of
galaxies and supercluster properties. 
We have obtained parameters useful to characterize the stellar
population of galaxies using empirical spectral synthesis of SDSS spectra
obtained with 
the STARLIGHT code of \citet{CidFernandesetal2005}.
This technique aims at reproducing an observed galaxy spectrum with a linear 
combination of simple stellar populations (SSPs), taking into account 
extinction and line broadening due to the stellar velocity dispersion.
It allows to estimate physical properties of galaxies, such as star 
formation and chemical enrichment histories and mean stellar ages and 
metallicities \citep[e.g.][]{CidFernandesetal2004,Tojeiroetal2007}. 

This paper is organized as follows. In Section \ref{data} we
present the
data analyzed here: the supercluster catalogue and the results of the
spectral synthesis of supercluster galaxies. In Section \ref{sc_prop} we discuss the
connection between stellar population parameters and supercluster properties,
as well as that between galaxy clusters and superclusters. Finally, in Section
\ref{conc_disc}, we summarize our main results and discuss their main
implications for our knowledge of large scale structures.  

\section{The data}
\label{data}

In this section we describe the supercluster
catalogue (CD11) and the supercluster parameters (morphology, richness, total luminosity).
 We also present the quantities associated to the stellar populations
(mean stellar ages and metallicities) and galaxy environment (local density),
describing how they were obtained.

We assume an standard $\Lambda$CDM cosmology, when necessary, with
$\Omega_m=0.3$, $\Omega_\Lambda=0.7$, and Hubble parameter $H_0 = 100$ h$^{-1}$ km
s$^{-1}$ Mpc$^{-1}$.

\subsection{Supercluster identification}
\label{sc_id}
The supercluster catalogue analyzed in this work is discussed in detail in CD11.
We identified superclusters by using a volume-limited sample of galaxies 
extracted from the main footprint of the Sloan Digital Sky Survey
\citep[SDSS/DR7,][]{Abazajianetal2009}, containing 120013 galaxies 
brighter than M$_r<$-21+5$\log$h in the redshift range 
0.040$\le$z$\le$0.115. Here we present a 
brief description of our approach, refering the reader to CD11 for more details.

We have adopted the density field method to find high density regions in the 
galaxy 
distribution. The luminosity density at a point $\mathbf{r}$ is given by
\begin{center}
\begin{equation}
D(\mathbf{r})=\sum_i K(|\mathbf{r-r_i}|,\sigma)L_i W_{i}(\mathbf{r_i}),
\label{eq_dens}
\end{equation}
\end{center}
where $K(r,\sigma)$ is the Epanechnikov kernel, $L_i$ is the luminosity of 
the i-th galaxy within the kernel radius, and $W_{i}$ is a
statistical weight taking into account the selection effects. The density 
field grid has cells with length $l_{cel}$=4~$h^{-1}$Mpc and the
kernel smoothing parameter is $\sigma$=8~$h^{-1}$Mpc. The luminosity density,
which is used to find superclusters,
is also useful for characterizing the local galaxy 
environment in investigations of stellar populations (see next Section).

To evaluate the influence of the threshold density on the supercluster 
identification, our analysis has been performed with two thresholds,
$D_1=3.0 \times D_0$ and 
$D_2=6.0 \times D_0$, where $D_0$ is the mean luminosity density. 
The first threshold maximizes the number of structures and the 
second is based on the \cite{Einastoetal2007a} criterion, assuming that the
largest supercluster has length $\sim$120$h^{-1}$Mpc. 

The morphology of superclusters was determined with Minkowski Functionals.
We have considered the parameters of planarity (K$_1$) and filamentarity (K$_2$) 
to classify superclusters as filaments (K$_1$/K$_2\leq$1.0) or
pancake-like objects (K$_1$/K$_2>$1.0)\citep{Sahnietal1998}. 
This ratio increases monotonically from filaments to pancakes.
The richness (R) and total luminosity ($L_{tot}$) have been measured for
all superclusters in our catalogue (see CD11 for more details). 

Table \ref{table_sc_results} shows some properties of the supercluster 
samples associated to each threshold. This table presents the number of
filamentary and pancake-like superclusters, the number of clusters (see section
\ref{cluster_prop_sc}) within
each morphological group and the number of galaxies for each 
threshold. Obviously, some of these galaxies are in both samples: 10953 of 
them. The total number of galaxies in this analysis is 50422.    

 We have verified that, 
qualitatively, the results of the next sections summarized in Figures 1 to 4 are
not affected by the choice of the threshold and, hereafter, all our results
will refer to $D_1$.

For each supercluster, structural (richness $R$, total luminosity 
$L_{tot}$) and morphological (the estimator $K_1/K_2$, obtained with Minkowski
functionals) parameters were computed (see CD11 for details), as well as the 
median density contrast, or overdensity $<D/D_0>$, 
considering all galaxies in each supercluster.

\begin{table}
\begin{center}
 \caption{Supercluster samples. For each threshold density we give the 
number of filaments (N$_{f}$), the number of pancakes (N$_{p}$), the number of 
clusters in filaments (N$_{cf}$), the number of clusters in pancakes (N$_{cp}$), and the number of galaxies (N$_{g}$).}
 \label{table_sc_results}
 \begin{tabular}{cccccc}
  \hline
threshold & N$_{f}$ & N$_{p}$ & N$_{cf}$ &
N$_{cp}$ & N$_{g}$
\\
  \hline
D$_1$ & 436 & 444 & 74 & 59 & 39469 \\
D$_2$ & 215 & 194 & 64 & 60 & 22817 \\
  \hline
 \end{tabular}
\end{center}
\end{table}

\subsection{Spectral synthesis}
\label{sec_spectral_synthesis}

The parameters related to the stellar populations of the galaxies in the
sample were obtained through an analysis of the galaxy spectra with 
the STARLIGHT code \citep{CidFernandesetal2005}, which makes use of
techniques of empirical population synthesis and evolutionary models to fit 
each observed spectrum with a combination of simple stellar populations (SSPs).

We adopted a spectral library with $N_{\star}$=150 components (SSPs) from
\cite{BC03}, with the initial mass function of \cite{Chabrier2003},
evolutionary tracks of Padova 1994 \citep{Alongietal1993,Girardietal1996} and 
the STELIB library \citep{LeBorgneetal2003}. The spectral library contains 
25 ages in the range 10$^6$ $\le$ t $\le$ 18$\times$10$^9$ years, 6 
metallicities in the range 0.0001 $\le$ Z $\le$ 0.05 and the extinction 
parameter (A$_V$) was constrained to the interval
-1.0 $\le$ $A_V$ $\le$ 4.0. The STARLIGHT code provides several
interesting parameters useful for this paper and we select for our analysis
the mean stellar ages and metallicities 
($<\log(t)>$ and $<Z>$) weighted by light and by mass.
\cite{CidFernandesetal2005}
defined the mean ages and metallicities weighted 
by the light vector $\left\{x \right\}$ as follows,
\begin{eqnarray}
 <\log(t)>_L&=&\sum_{i=1}^{N_{\star}} x_i \log(t_i) \\
 <Z>_L&=&\sum_{i=1}^{N_{\star}} x_i Z_i. 
\end{eqnarray}
In a similar way, the mean age and metallicity weighted by mass are
defined using the mass vector ($\left\{\mu\right\}$). The spectral synthesis was
carried out for 50422 galaxies, and the relevant stellar population properties 
were calculated for each object. 

Using the SDSS spectral classification, we identified
420 objects in our sample as QSOs. Seyfert I galaxies frequently present a 
spectral continuum hard to be fitted by stellar populations. We then carried 
out a visual inspection of the spectral fitting of these objects,
finding  that spectral synthesis produced bad fittings for 259 objects,
which were consequently excluded from our stellar population analysis. 

For each supercluster in our sample we computed the median values
of the stellar ages and metallicities of their galaxies. These median values
are the stellar population parameters that we will ascribe to each supercluster.
It is worth mentioning that these parameters were obtained for galaxies 
brighter than M$_r<$-21+5$\log$h and, hence, they are not representative of the 
whole stellar population. However we use the same luminosity threshold for
all structures and consequently the properties of bright galaxies can be
compared between superclusters with different properties.

\begin{table}
\begin{center}
\caption{Spearman rank-order correlation coefficient ($r_s$) and the 
probability of absence of correlation ($P(H_0$)) for the relation between
supercluster parameters and their stellar population properties. The
ages and metallicities weighted by mass and light are represented with the
subscripts M and L, respectively.}
\label{spearman}
\begin{tabular}{lrr}
\hline
Variables & $r_s$ & $P(H_0)$ \\ \hline
K$_1$/K$_2$ - $<$log(t)$>_M$ & -0.051 & 0.130\\
K$_1$/K$_2$ - $<$log(t)$>_L$ & -0.082 & 0.015\\
K$_1$/K$_2$ - $<$Z$>_M$ & 0.029 & 0.392\\
K$_1$/K$_2$ - $<$Z$>_L$ & 0.030 & 0.370\\
R - $<$log(t)$>_M$ & 0.142 & $<10^{-3}$\\
R - $<$log(t)$>_L$ & 0.142 & $<10^{-3}$\\
R - $<Z>_M$ & -0.042 & 0.218\\
R - $<$Z$>_L$ & -0.074 & 0.027\\
L$_{tot}$ - $<$log(t)$>_M$ & 0.135 & $<10^{-3}$\\
L$_{tot}$ - $<$log(t)$>_L$ & 0.139 & $<10^{-3}$\\
L$_{tot}$ - $<$Z$>_M$ & -0.019 & 0.574\\
L$_{tot}$ - $<$Z$>_L$ & -0.039 & 0.244\\
$<$D/D$_0$$>$ - $<$log(t)$>_M$ & 0.117 & $<10^{-3}$\\
$<$D/D$_0$$>$ - $<$log(t)$>_L$ & 0.129 & $<10^{-3}$\\
$<$D/D$_0$$>$ - $<$Z$>_M$ & -0.007 & 0.844\\
$<$D/D$_0$$>$ - $<$Z$>_L$ & -0.001 & 0.983\\ \hline
\end{tabular}
\end{center}
\end{table}

\section{Properties of Superclusters}
\label{sc_prop} 

In this section we first study the relation between stellar
population parameters and other supercluster properties, then
we address the relation between galaxy clusters and superclusters.

\subsection{Stellar populations}
\label{global_prop}

To evaluate the influence of supercluster properties on the stellar
population of galaxies, we have investigated the correlations between 
supercluster richness ($R$), total luminosity ($L_{tot}$), median galaxy
overdensity ($<D/D_0>$) and 
morphology ($K_1$/$K_2$), and the representative values of mean ages and
metallicities 
of galaxies belonging to each supercluster. 
The strength of each correlation is evaluated with
the nonparametric Spearman rank-order correlation coefficient \citep{press}.
The parameter r$_s$ represents the rank correlation, a number between -1
(anti-correlation) and +1 (correlation),
and  $P(H_0)$ is the null-hypothesis probability of absence of correlation.
Table \ref{spearman} summarizes the results of this analysis. 

The first result to note is that stellar populations do not depend at all
on supercluster morphology, as indicated by the large values of  $P(H_0)$
in the relations with the morphological indicator $K_1/K_2$. This result
indicates that the morphology does not have a significative influence on
the stellar population of galaxies. 

\begin{figure}
\includegraphics[scale=0.7]{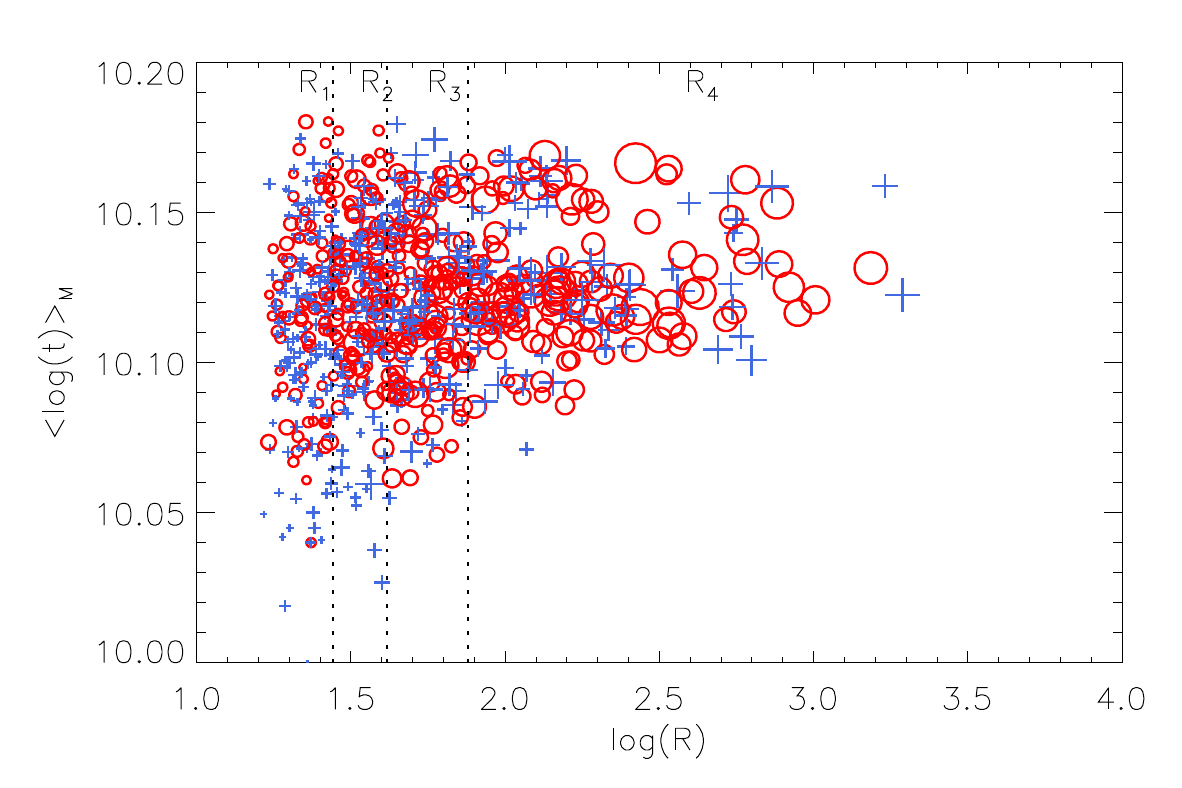} 
\caption{The median ages weighted by mass ($<log(t)>_M$)
of superclusters as function of
their richness ($R$). The size of the symbols is proportional to the median
supercluster density contrast($<D/D_0>$). 
Superclusters classified as filaments and
pancakes are shown as circles (red) and crosses (blue), respectively. The
vertical dashed lines represent the median and quartiles of richness, which
we use to define the samples $R_1$, $R_2$, $R_3$ and $R_4$ (see text). }
\label{R_logt_sc}
\end{figure}

There are significant correlations (i.e., with low values of $P(H_0)$) 
between supercluster richness, total luminosity and mean density with mean ages:
richer, more luminous and denser superclusters tend to harbour older
stellar populations. These correlations, however, are weak, as indicated
by the small (absolute) values of $r_s$. Figure \ref{R_logt_sc} shows
the median value of $<\log(t)>_M$ as a function of the supercluster richness. 
Similar results are obtained for the relations between ages and total
luminosities and median overdensities.

There are no significative relations between the supercluster properties and the
mean metallicities of their galaxies. It is interesting to notice in Table
\ref{spearman} that age and metallicity correlation coefficients have 
always opposite signs. This may be an indication of the age-metallicity
degeneracy and, since our results are more sensitive to ages than to
metallicities, it is possible that a trend with metallicity is weakened and 
at the same time the trend with age is strengthened due to this degeneracy.

To evaluate the significance of the absence of rich, luminous and high density 
superclusters with ages as young as those found for low-richness structures
(Figure \ref{R_logt_sc}), and verify whether this is only an observational
effect due to the relatively small number of high-density structures, we
divided the galaxies in four bins of richness (and also of luminosity
and of overdensity). We computed the median and quartiles of the distribution
and defined the four richness bins ($R_1$, $R_2$, $R_3$ and
$R_4$) by the quartiles of the distribution. The Kolmogorov-Smirnov test was
then applied to compare the samples of superclusters in
different richness bins. We verified that there are significant
differences in the age distributions of galaxies
in the first two richness quartiles with respect to galaxies in the fourth 
quartile. The same is true if we analyze the behaviour of age distribution
in bins of $L_{tot}$ and $<D/D_0>$.
These results indicate that the age distribution of
superclusters in different richness/luminosity/density classes can be 
considered statistically distinct (mostly by comparing samples of galaxies
in the bins 1 and 4). Thus the absence of rich, luminous and denser
superclusters with young stellar populations seems to be indeed an environmental
effect.

\begin{figure*}
\includegraphics[scale=0.42]{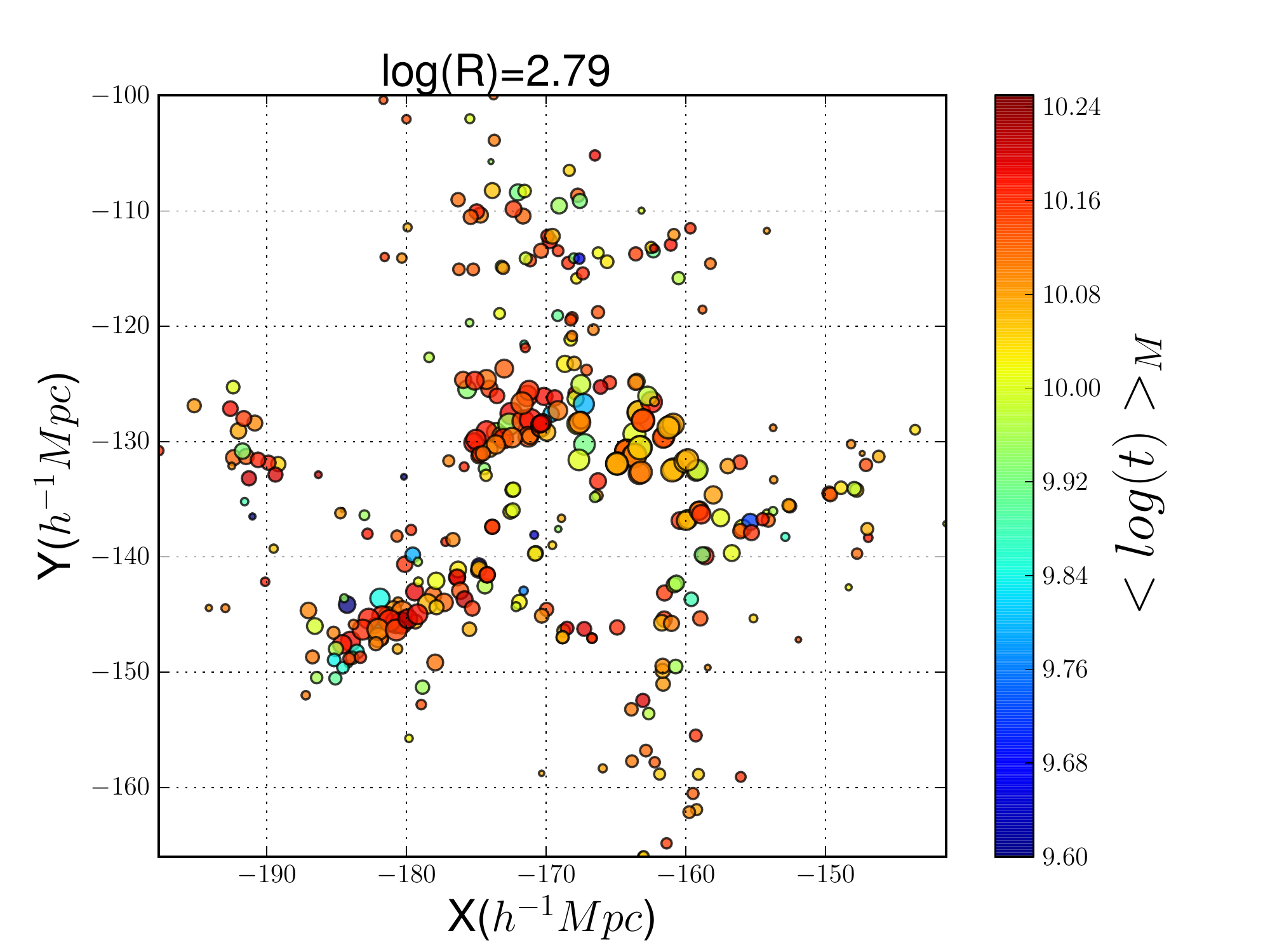} 
\includegraphics[scale=0.42]{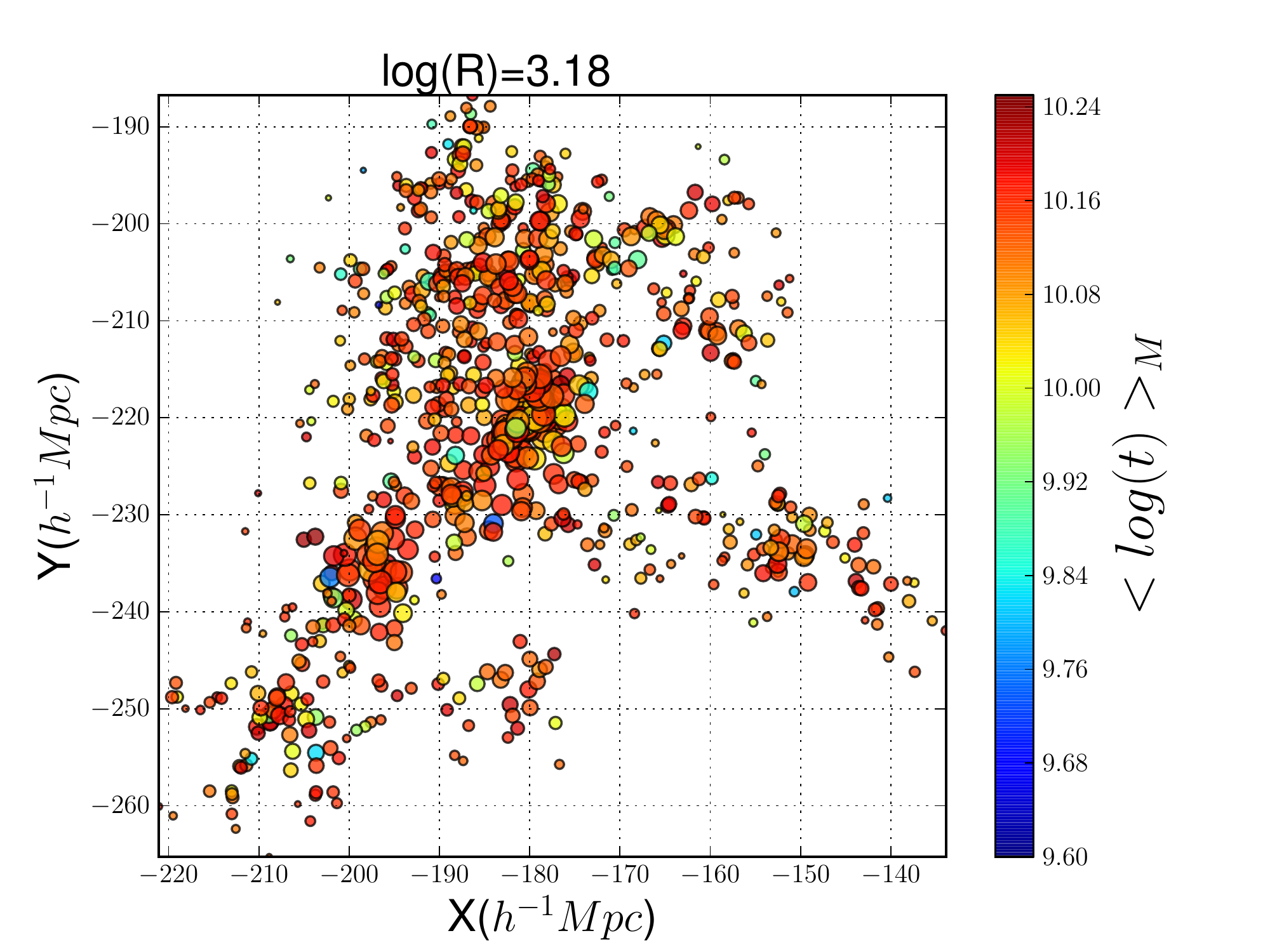} 

\includegraphics[scale=0.42]{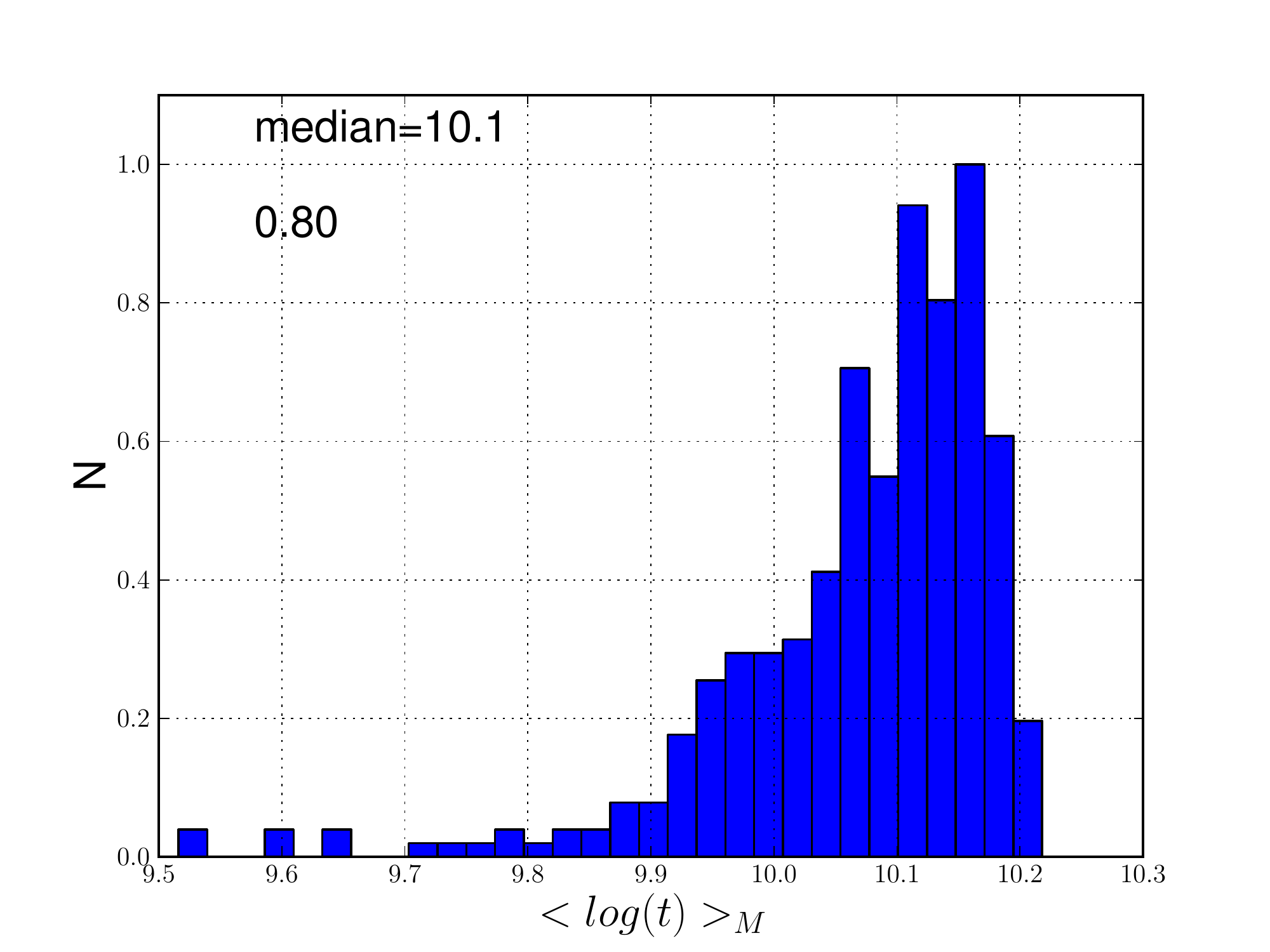} 
\includegraphics[scale=0.42]{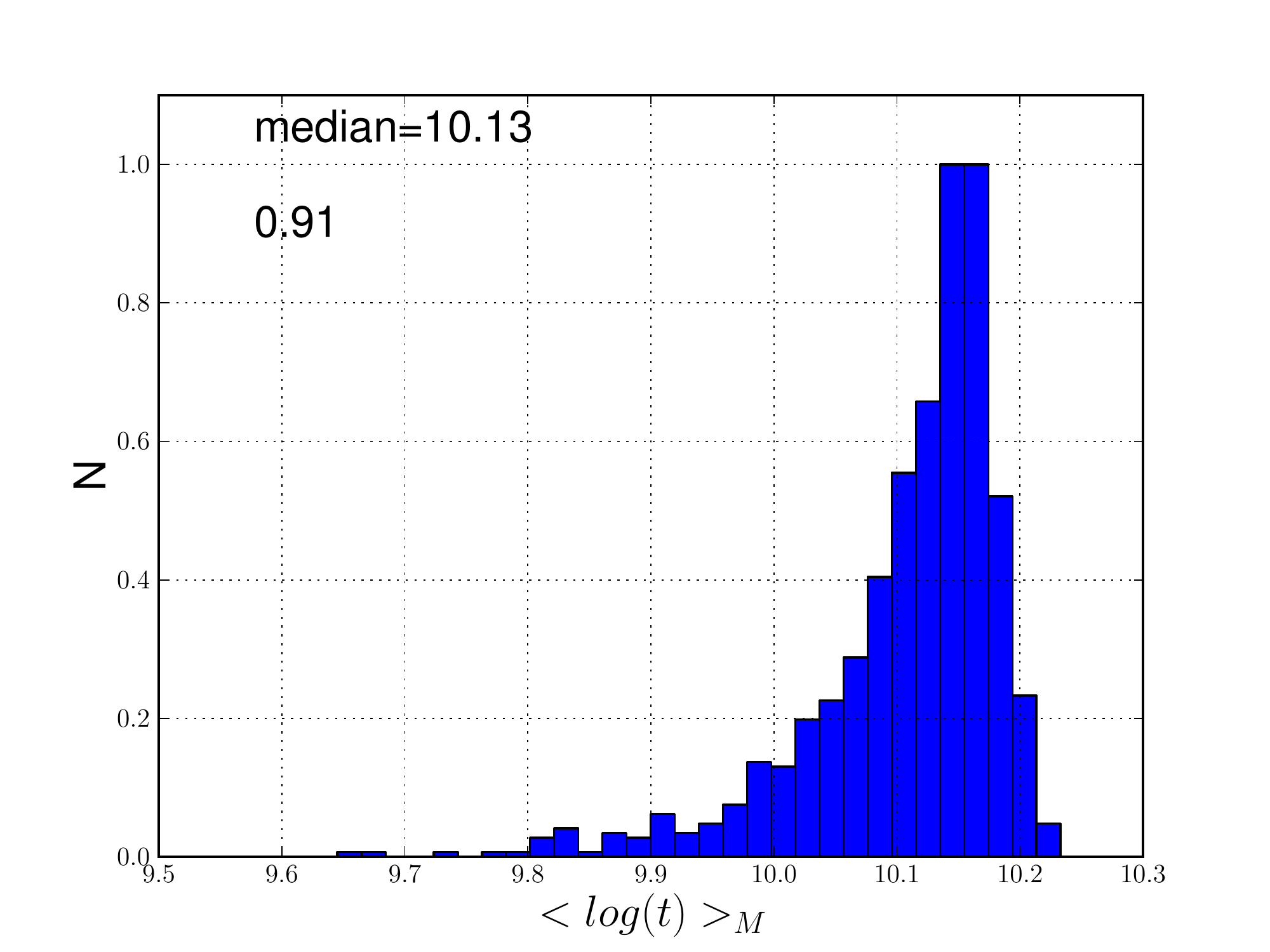} 

\caption{Top: Spatial distribution of two superclusters
from
our sample with different richness. The radii
of the circles are proportional to the local
luminosity density of each galaxy 
 and the galaxy ages weighted by mass ($<log(t)>_M$) follow
the gray (color) scale. Bottom : Histogram of $<log(t)>_M$ of
galaxies which belongs to the superclusters, showing the median value
and the fraction of galaxies with $<log(t)>_M$ greater than 10 Gyrs.}
\label{dens_gal_r_sc}
\end{figure*}

Richer superclusters present higher density regions than less rich structures.
Invoking the morphology-density relation, we may expect that 
richer superclusters present, on average, galaxies with older
stellar populations. High density regions strongly affect
star-forming galaxies, quenching the star formation by several
processes, such as starvation, ram pressure, etc. As a consequence, an older
stellar population dominates in these regions after roughly one 
1Gyr after the arrival of (infalling) star forming galaxies. 
Since high-density structures are much more common in high-richness objects,
it is reasonable to expect that these structures present, on average, an older 
stellar population.Figure~\ref{dens_gal_r_sc} shows the galaxy
distribution of two superclusters with different richness.
The top figures show filaments of galaxies linking the high
density
regions, which are mostly populated by galaxies with old stellar
populations, while the fraction of young stellar population increases in
low density regions. The poorer supercluster shown in this figure does not have
regions with densities as high
as those found in the richer structure, and, consequently has
a significantly younger stellar population. The bottom figures show
histograms of $<log(t)>_M$
of galaxies for these superclusters. The median values and the fractions of
galaxies with age higher than 10 Gyrs also indicate different stellar
populations. Thus the influence of the
environment on stellar populations will be distinct for
each supercluster, according to its richness and consequently total
luminosity. 


\subsection{Clusters in Superclusters}
\label{cluster_prop_sc}

Superclusters are, actually, formed by groups and clusters of galaxies
and it is then interesting to verify whether the properties of, for
example, galaxy clusters, are related to the properties of the supercluster 
which they inhabit.

For this investigation we have adopted the cluster
catalogue of \cite{Wenetal2009}, which contains 39668 objects from 
SDSS/DR6 in the redshift range 0.05$<$z$<$0.6. Clusters are
identified as supercluster members if their centres are 
within a supercluster. Table \ref{table_sc_results} presents the
number of clusters identified in superclusters classified as filaments
and pancakes. The cluster catalogue contains several cluster properties, such as
richness (R$_{clus}$) and r-band luminosity (L$_{clus}$), which are useful
proxies of cluster mass \citep[e.g.,][]{Wenetal2010}. 

\begin{figure}
\includegraphics[scale=0.7]{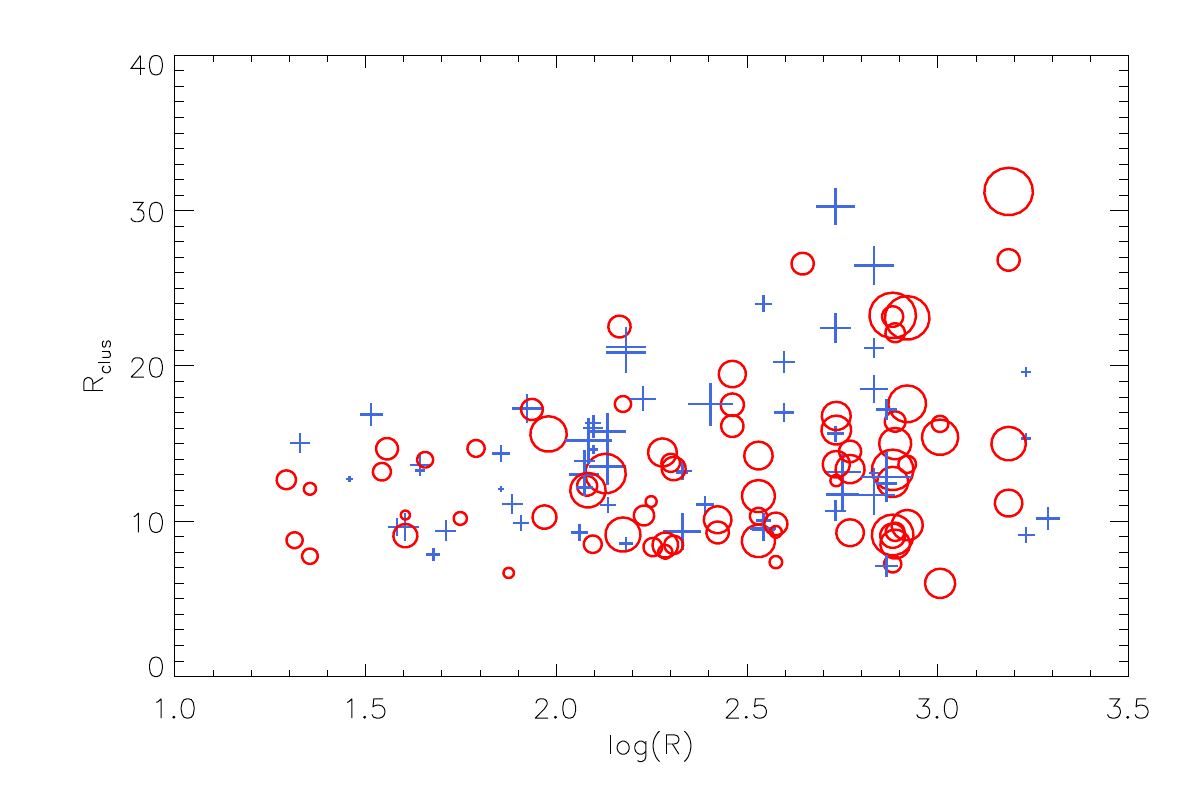} 
\caption{The richness of clusters ($R_{clus}$) as function of
supercluster richness ($R$). The size of the symbols is proportional to the
density contrast($<D/D_0>$). 
Clusters in superclusters classified as filaments and
pancakes are shown in circles (red) and crosses (blue), respectively.}
\label{R_clus_R_sc}
\end{figure}

We verified that galaxy cluster properties are not related to the 
supercluster morphology. Indeed, the distribution of cluster richness
and luminosity is statistically the same for filaments and pancakes. We have
also
verified that richer and denser clusters are preferentially located in richer
superclusters, confirming the \cite{Einastoetal2012} results, as shown in
figure \ref{R_clus_R_sc}.

\begin{figure*}
\includegraphics[scale=0.8]{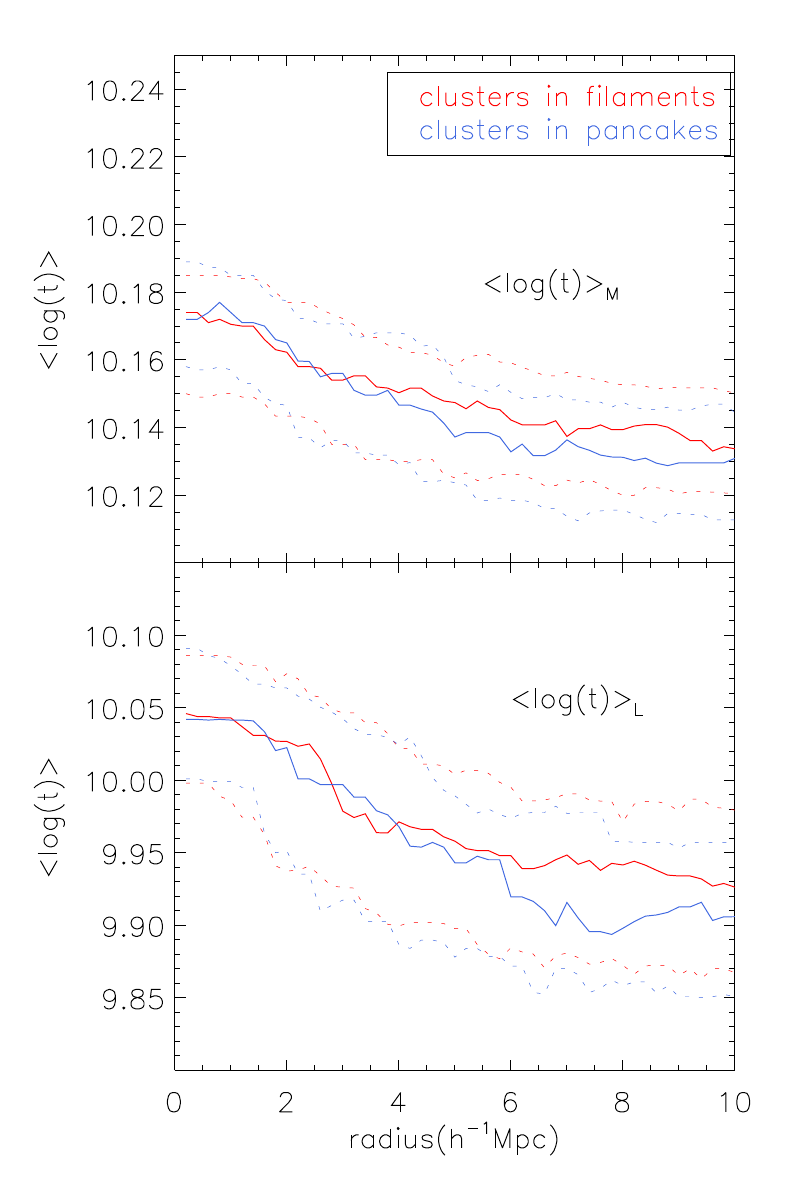} 
\includegraphics[scale=0.8]{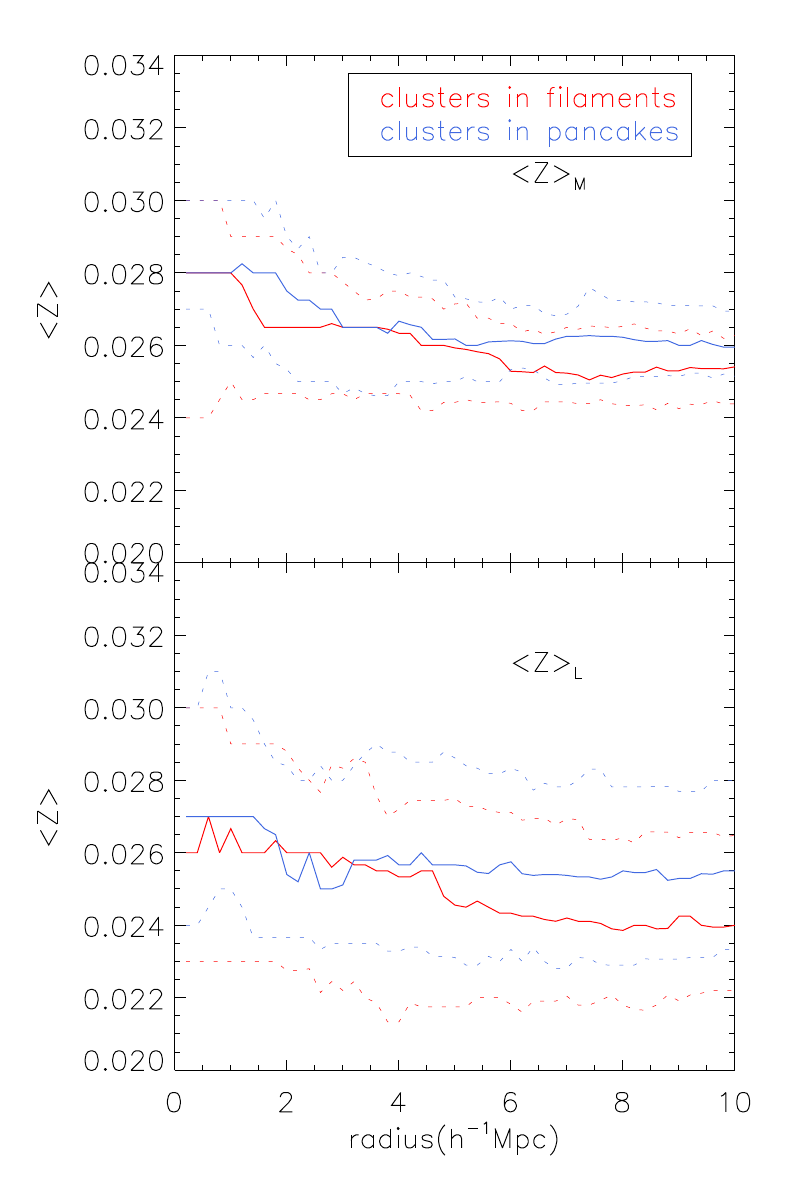} 
\caption{Left: median profiles of ages weighted by mass and
light of clusters in
filaments (red) and pancakes (blue). Right: median metallicity profiles of
clusters. The parameters weighted by mass and light are presented in the top
and bottom panels, respectively. The thick lines represent the median values and
the dashed lines
represent the quartiles for each radius.}
\label{profile_clusters_logt}
\end{figure*}

We have also evaluated the effect of environment on galaxies in clusters and
in their outskirts. For this, we calculated,  for each cluster, 
mean values of ages and metallicities of galaxies inside spherical shells 
with radii varying from 0.1h$^{-1}$Mpc to 10h$^{-1}$Mpc for each cluster. 
At small radii, only cluster galaxies contribute to these means, and the
results indicate the prevalence of an older and more metallic population that 
those found at larger radii: as the radii increase, the mean values of ages 
and metallicities decrease, due to an increasing contribution of 
late-type galaxies,  which present a higher fraction of younger and less
enriched stellar populations. 
Figure \ref{profile_clusters_logt} shows the median
profiles of ages and metallicities weighted by light and mass for clusters
in filaments and pancakes. The trends indicate that clusters affect the
parameters probed by spectral synthesis up to a radius of 
$\sim$8h$^{-1}$Mpc. These profiles, however, do not present any
significant difference between clusters in filaments or in pancakes, nor
depend on supercluster richness, suggesting that the environmental effect on 
galaxies is driven fundamentally by clusters more than by the superclusters.
\section{Summary and Discussion}
\label{conc_disc}

In this work we have analyzed some properties of supercluster of galaxies,
making use of the supercluster catalogue described in CD11 which
considers
galaxies brighter than M$_r<$-21+5$\log$ h. The results focused on
connections 
between supercluster properties and their stellar populations. 
We have also examined some
links between galaxy clusters and superclusters.

Our main results can be summarized as follows:
\begin{enumerate}
\item the stellar populations of galaxies in superclusters are not affected by
supercluster morphology;

\item richer and more luminous superclusters have higher density regions;

\item richer and more luminous superclusters have, on average, older stellar
populations;

\item we do not find any significative correlation between supercluster
properties and the metallicity of the stellar populations; we suggest that
this may be an artifact produced by the age-metallicity degeneracy;

\item rich clusters tend to inhabit rich superclusters;

\item the behavior of stellar population parameters around clusters
seem to be independent of supercluster richness and morphology;

\end{enumerate}

It is worth stressing that most of the correlations discussed in this paper
are not strong but the trends are statistically significant.

The supercluster morphology is a tracer of the processes of structure
formation and evolution \citep[][]{Shandarinetal2004,Einastoetal2011a,
Einastoetal2011b}, and there is
evidence that filamentary structures tend to be richer, larger and more 
luminous than pancakes \cite[e.g., CD11, ][]{Einastoetal2011a},
suggesting an evolutionary track where pancakes evolve to filaments. 
We do not find, however, any significant correlation
between supercluster morphology and the properties of stellar populations.
On the other side, the influence of the environment on galaxy properties
is well described by various authors 
\citep[c.f.][]{Dressler1980, Mateusetal2007}. Our results might be indicating
that the supercluster morphology is not strongly correlated with the inner
environment of these structures, at least to the point of affecting the
stellar populations of their galaxies. So, although our results are all
consistent with the idea that the environment affects galaxy properties,
this influence should operate on scales of groups and clusters, more than in
the scale of superclusters.

\section*{Acknowledgments}

MVCD thanks FAPESP and CAPES scholarships that allowed him to develop this
project. LSJ acknowledges the support of FAPESP and CNPq to his work. 
This work is also supported by the CAPES/COFECUB project 711/11.

\label{lastpage}


\begin{thebibliography}{99}

\bibitem[Abazajian et al.(2009)]{Abazajianetal2009}
{Abazajian}, K.~N., {Adelman-McCarthy}, J.~K., {Ag{\"u}eros}, M.~A. et
al., 2009, \apjs, 182, 543

\bibitem[Abell et al.(1989)]{Abelletal1989}
{Abell}, G.~O., {Corwin}, Jr., H.~G. and {Olowin}, R.~P., 1989, \apjs, 70, 1

\bibitem[Alongi et al.(1993)]{Alongietal1993}
{Alongi} M.,  {Bertelli} G.,  {Bressan} A. et al., 1993, A\&AS, 97, 851

\bibitem[Bruzual \& Charlot(2003)]{BC03}
{Bruzual}, G. and {Charlot}, S., 2003, \mnras, 344, 1000

\bibitem[Carollo et al.(2012)]{Carolloetal2012}
{Carollo}, C.~M. and {Cibinel}, A. and {Lilly}, S.~J. et al., 2012,
arXiv:1206.5807

\bibitem[Chabrier(2003)]{Chabrier2003}
Chabrier G., 2003, PASP, 115, 763

\bibitem[Cid Fernandes et al.(2004)]{CidFernandesetal2004}
{Cid Fernandes}, R., {Gu}, Q., {Melnick}, J. et al., 2004, \mnras, 355, 273

\bibitem[Cid Fernandes et al.(2005)]{CidFernandesetal2005}
{Cid Fernandes}, R., {Mateus}, A., {Sodr{\'e}}, L.,{Stasi{\'n}ska}, G.,
{Gomes}, J.~M., 2005, \mnras, 358, 363

\bibitem[Colless et al.(2001)]{Collessetal2001}
{Colless}, M., {Dalton}, G., {Maddox}, S. et al., 2001, \mnras, 328, 1039

\bibitem[Costa-Duarte et al.(2011)]{Costa-Duarteetal2011}
Costa-Duarte, M. V., Sodr\'e Jr. L., Durret F., 2011, MNRAS, 411, 1716 (CD11)

\bibitem[Croton et al.(2005)]{Crotonetal2005}
{Croton}, D.~J. and {Farrar}, G.~R. and {Norberg}, P. et al., 2005, \apj, 356,
1155

\bibitem[Croton et al.(2006)]{Crotonetal2006}
{Croton}, D.~J., {Springel}, V., {White} S.D.M., et al., 2006, MNRAS, 365, 11

\bibitem[Davis \& Geller(1976)]{DavisGeller1976}
{Davis}, M. and {Geller},
M.~J., 1976, \apj, 208, 13

\bibitem[Dressler(1980)]{Dressler1980}
Dressler, A., 1980, \apj, 236, 351

\bibitem[Einasto et al.(1994)]{Einastoetal1994}
{Einasto}, M., {Einasto}, J., {Tago}, E., {Dalton}, G.~B., {Andernach}, H.,
1994, \mnras, 269, 301

\bibitem[Einasto et al.(2007a)]{Einastoetal2007a}
{Einasto} J.,  {Einasto} M.,  {Tago} E. et al.,
2007a,
A \& A, 462, 811

\bibitem[Einasto et al.(2007b)]{Einastoetal2007b}
{Einasto}, M. and {Einasto}, J. and {Tago}, E. et al., 2007b, \aap, 464, 815


\bibitem[Einasto et al.(2011a)]{Einastoetal2011a}
{Einasto}, M. and {Liivam{\"a}gi}, L.~J. and {Tago}, E. et al., 2011a, \aap,
532, A5

\bibitem[Einasto et al.(2011b)]{Einastoetal2011b}
{Einasto}, M., {Liivam{\"a}gi}, L.~J., {Saar}, E. et al., 2011b, \aap, 535, A36

\bibitem[Einasto et al.(2012)]{Einastoetal2012}
{Einasto}, M., {Liivam{\"a}gi}, L.~J., {Tempel}, E. et al., 2012, \aap,
542, A36

\bibitem[Gal \& Rubin(2004)]{GalRubin2004}
{Gal}, R.~R. and {Lubin}, L.~M., 2004, \apjl, 607, L1

\bibitem[Girardi et al.(1996)]{Girardietal1996}
{Girardi} L., {Bressan} A., {Chiosi} C., {Bertelli} G., {Nasi} E., 
1996, A\&AS, 117, 113

\bibitem[Gott et al.(2005)]{Gottetal2005}
{Gott}, III, J.~R. and {Juri{\'c}}, M. and {Schlegel}, D. et al., 2005, \apj,
624, 463

\bibitem[Kuiper et al.(2012)]{Kuiperetal2012}
{Kuiper}, E., {Venemans}, B.~P., {Hatch}, N.~A., {Miley}, G.~K.,
{Rottgering}, H.~J.~A., 2012, \mnras, 425, 801

\bibitem[Koester et al.(2007)]{Koesteretal2007}
{Koester}, B.~P., {McKay}, T.~A., {Annis}, J. et al., 2007, \apj,
660, 239

\bibitem[Le Borgne et al.(2003)]{LeBorgneetal2003}
{Le Borgne} J.-F.,  {Bruzual} G.,  {Pell{\'o}} R. et al., 2003, A\&A, 402, 433

\bibitem[Lietzen et al.(2012)]{Lietzenetal2012}
{Lietzen}, H. and {Tempel}, E. and {Hein{\"a}m{\"a}ki}, P. et al., 2012,
arXiv:1207.7070

\bibitem[Lilly et al.(2007)]{Lillyetal2007}
{Lilly}, S.~J., {Le F{\`e}vre}, O., {Renzini}, A. et al., 2007, \apjs, 172, 70

\bibitem[Luparello et al.(2011)]{Luparelloetal2011}
{Luparello}, H., {Lares}, M., {Lambas}, D.~G., {Padilla},
N., 2011, \mnras, 415, 964

\bibitem[Mateus et al.(2007)]{Mateusetal2007}
{Mateus}, A., {Sodr{\'e}}, L., {Cid Fernandes}, R., {Stasi{\'n}ska}, G.,
2007, MNRAS, 374, 1457

\bibitem[Mo et al.(1992)]{Moetal1992}
{Mo}, H.~J., {Einasto}, M., {Xia}, X.~Y., {Deng}, Z.~G.,
1992, \mnras, 255, 382

\bibitem[Park et al.(2007)]{Parketal2007}
{Park}, C. and {Choi}, Y.-Y. and {Vogeley}, M.~S. and {Gott}, III, J.~R.
and {Blanton}, M.~R. and {SDSS Collaboration},
2007, \apj, 658, 898

\bibitem[Press et al.(1992)]{press}
{Press}, W. H., {Teukolsky}, S. A., {Vetterling}, W. T., {Flannery}, B. P.,
1992, Numerical Recipes, 2nd. edition, CUP

\bibitem[Popesso et al.(2009)]{Popessoetal2009}
{Popesso}, P., {Dickinson}, M., {Nonino}, M. et al., 2009, \aap, 494, 443

\bibitem[Sahni, Sathyaprakash \& Shandarin(1998)]{Sahnietal1998}
{Sahni} V.,  {Sathyaprakash} B.~S., {Shandarin} S.~F., 1998, \apjl, 495, L5

\bibitem[Shandarin, Sheth \& Sahni(2004)]{Shandarinetal2004}
{Shandarin}, S.~F. and {Sheth}, J.~V. and {Sahni}, V., 2004, \mnras, 353, 162

\bibitem[Shectman et al.(1996)]{Schectmanetla1996}
{Shectman} S.A., {Landy} S.D., {Oemler} A. et al., 1996, \apj, 470, 172S

\bibitem[Tanaka et al.(2001)]{Tanakaetal2001}
{Tanaka}, I. and {Yamada}, T. and {Turner}, E.~L. and {Suto}, Y.,
2001, \apj, 547, 521

\bibitem[Tojeiro et al.(2007)]{Tojeiroetal2007}
{Tojeiro}, R., {Heavens}, A.~F., {Jimenez}, R., {Panter}, B.,
2007, \mnras, 381, 1252

\bibitem[Wen et al.(2009)]{Wenetal2009}
{Wen}, Z.~L., {Han}, J.~L., {Liu}, F.~S., 2009, ApJS, 183, 197

\bibitem[Wen et al.(2010)]{Wenetal2010}
{Wen}, Z.~L., {Han}, J.~L., {Liu}, F.~S., 2010, MNRAS, 407, 533

\bibitem[Whitmore, Gilmore \& Jones(1993)]{Whitmoreetal1993}
{Whitmore}, B.~C., {Gilmore}, D.~M., {Jones}, C., 1993, \apj, 407, 489

\bibitem[Zucca et al.(1993)]{Zuccaetal1993}
{Zucca}, E., {Zamorani}, G., {Scaramella}, R., {Vettolani},
1993, \apj, 407, 470

\end{thebibliography}
\end{document}